\documentclass[a4paper,11pt]{article}
\usepackage{amssymb}
\usepackage{amsmath}
\usepackage{amsmath,amssymb,amssymb,amsfonts}
\usepackage{graphicx,,epsfig}
\usepackage{slashed,xspace}
\usepackage{mathrsfs}
\usepackage{dsfont}
\usepackage{wasysym}

\newcommand{\be}{\begin{equation}}
\newcommand{\ee}{\end{equation}}
\newcommand{\bea}{\begin{eqnarray}}
\newcommand{\eea}{\end{eqnarray}}
\newcommand{\ba}{\begin{array}}
\newcommand{\ea}{\end{array}}
\newcommand{\bit}{\begin{itemize}}
\newcommand{\eit}{\end{itemize}}
\newcommand{\ben}{\begin{enumerate}}
\newcommand{\een}{\end{enumerate}}

\makeatletter \@addtoreset{equation}{section} \makeatother

\begin{document}

\begin{titlepage}

    \thispagestyle{empty}
    \begin{flushright}
         %\hfill{LNF-09-...}\\
         \hfill{CERN-PH-TH/2009-200} \\
        %\hfill{UCLA/09/TEP/...}\\
        \hfill{SU-ITP-09/47}\\
    \end{flushright}

    %\vspace{12pt}
    \begin{center}
        { \huge{\bf Maurer-Cartan Equations\\\vspace{5pt}and Black Hole Superpotentials\\\vspace{10pt}in $\mathcal{N}=8$ Supergravity}}

        \vspace{18pt}

        {\large{\bf Sergio Ferrara$^{\diamondsuit,\spadesuit,\flat}$, Alessio Marrani$^{\heartsuit}$\\and \ Emanuele Orazi$^{\clubsuit}$}}

        \vspace{15pt}

       {$\diamondsuit$ \it Theory Division - CERN,\\
       CH 1211, Geneva 23, Switzerland\\
       \texttt{sergio.ferrara@cern.ch}}

\vspace{5pt}

        {$\spadesuit$ \it INFN - LNF, \\
         Via Enrico Fermi 40, I-00044 Frascati, Italy}

         \vspace{5pt}

         {$\flat$ \it Department of Physics and Astronomy,\\
       University of California, Los Angeles, CA USA}

        \vspace{5pt}

        {$\heartsuit$ \it Stanford Institute for Theoretical Physics\\
        %Department of Physics, 382 Via Pueblo Mall, Varian Lab,\\
        Stanford University, Stanford, CA 94305-4060, USA\\
        \texttt{marrani@lnf.infn.it}}

        \vspace{5pt}

        {$\clubsuit$ \it Dipartimento di Fisica, Politecnico di Torino,\\
        Corso Duca degli Abruzzi 24, I-10129 Turin, Italy\\
        \texttt{emanuele.orazi@polito.it}}

        %\vspace{15pt}
        %\noindent \textit{Contribution to the Proceedings of the\\School on Attractor Mechanism 2007 (SAM2007),\\June 18--22 2007, INFN--LNF, Frascati, Italy}
\end{center}

\vspace{15pt}

\begin{abstract}
We retrieve the non-BPS extremal black hole superpotential of
$\mathcal{N}=8$, $d=4$
supergravity by using the Maurer-Cartan equations of the symmetric space $%
\frac{E_{7\left( 7\right) }}{SU\left( 8\right) }$. This
superpotential was recently obtained with different $3$- and
$4$-dimensional techniques. The present derivation is independent on the
reduction to $d=3$.
\end{abstract}

\end{titlepage}

\section{\label{Intro}Introduction}

Recently, much progress has been obtained in the description of BPS and
non-BPS \textit{extremal} black hole (BH) flows in $\mathcal{N}\geqslant 2$
supergravities in $d=4$ space-time dimensions \cite
{Cer-Dal-fake,ADOT-1,stu-unveiled,ADOT-2,CDFY-1,BMP-1,CDFY-2} (see also
Sect. 2 of \cite{Gnecchi-1}). In particular, for all theories whose
non-linear scalar sigma model is a symmetric space\footnote{%
Note that this is always the case for $\mathcal{N}\geqslant 3$, $d=4$
theories.}, \textit{``superpotentials'' }$W$'s exist for all BPS and non-BPS
branches, thus yielding that the corresponding radial flow equations are of
first order. Namely, the warp factor $U$ of the extremal BH metric and the
scalar field trajectories respectively read \cite{Cer-Dal-fake}:
\begin{eqnarray}
\dot{U} &=&-e^{U}W; \\
\dot{\phi}^{i} &=&-2e^{U}g^{ij}\partial _{j}W,  \label{U-1}
\end{eqnarray}
where $W$ is related to the effective BH potential
\begin{equation}
V_{BH}\equiv \frac{1}{2}Z_{AB}\overline{Z}^{AB}+Z_{I}\overline{Z}^{I}
\label{V_BH-gen}
\end{equation}
through
\begin{equation}
V_{BH}=W^{2}+2g^{ij}\partial _{i}W\partial _{j}W=W^{2}+2g^{ij}\nabla
_{i}W\nabla _{j}W.  \label{V_BH-W}
\end{equation}
Here $Z^{I}$ denote the matter charges (absent \textit{e.g.} in $\mathcal{N}%
=8$ supergravity), and $Z_{AB}=-Z_{BA}$ is the central charge matrix,
entering the supersymmetry algebra as follows:
\begin{equation}
\left\{ \mathcal{Q}_{A}^{\alpha },\mathcal{Q}_{B}^{\beta }\right\} =\epsilon
^{\alpha \beta }Z_{AB}\left( \phi _{\infty },Q\right) .
\end{equation}
Moreover, Eq. (\ref{U-1}) implies that attractor points
\begin{equation}
\dot{\phi}^{i}=0
\end{equation}
correspond to critical points of $W$ itself:
\begin{equation}
\partial _{i}W=0.
\end{equation}

For BPS BHs
\begin{equation}
W\left( \phi ,Q\right) =\left| z_{I}\right| _{\max }\left( \phi ,Q\right) ,
\label{W-BPS-gen}
\end{equation}
where $Q$ is the symplectic charge vector, and $\left| z_{I}\right| _{\max }$
is the highest absolute value of the \textit{skew-eigenvalues} $z_{I}$'s of $%
Z_{AB}$. Furthermore, the \textit{ADM mass} $M_{ADM}$ \cite{ADM} is related
to $W$ through ($r$ denotes the radial coordinate throughout)
\begin{equation}
M_{ADM}^{2}=\lim_{r\rightarrow \infty }W^{2}.  \label{ADM}
\end{equation}
The Bekenstein-Hawking entropy-area formula \cite{BH1} exploits as follows:
\begin{equation}
\frac{S_{BH}\left( Q\right) }{\pi }=\frac{A_{H}}{4\pi }=\lim_{r\rightarrow
r_{H}^{+}}W^{2}=\left. W^{2}\right| _{\partial W=0}=W^{2}\left( \phi
_{H}\left( Q\right) ,Q\right) ,
\end{equation}
where $r_{H}$ and $A_{H}$ respectively stand for the radius and the area of
the event horizon of the considered extremal BH, and $\phi _{H}\left(
Q\right) $ denotes the set of scalar fields at the horizon, stabilized in
terms of the charges $Q$.

Explicit ways of constructing $W$ has been given in \cite
{CDFY-1,BMP-1,CDFY-2} by using different methods, e.g. based on the $%
\mathcal{N}=2$ $stu$ model \cite{CDFY-1,CDFY-2} or on three-dimensional
techniques \cite{BMP-1}. All these exploit the fact, as generally proven in
\cite{ADOT-2}, that
\begin{equation}
W=W\left( i_{n}\left( \phi ,Q\right) \right) ,
\end{equation}
where $i_{n}\left( \phi ,Q\right) $'s ($n=1,...,5$) are duality invariant
combinations of the scalars $\phi ^{i}$'s and of charges $Q$ \cite
{CFMZ-1,CDFY-1}. A polynomial in $i_{n}$'s gives the (unique)
scalar-independent duality invariant $\mathcal{I}\left( Q\right) $ \cite
{K-Kol,ADF-U-duality-d=4,CFMZ-1}. In the $\mathcal{N}=2$ case, it reads \cite
{CFMZ-1,CDFY-1}:
\begin{equation}
\mathcal{I}=\left( i_{1}-i_{2}\right) ^{2}+4i_{4}-i_{5}.
\end{equation}
It is worth remarking that in the considered framework the symplectic vector
of charges $Q$ must belong to a \textit{non-degenerate} (\textit{i.e.} with $%
\mathcal{I}\neq 0$) orbit of the $U$-duality group \cite
{BFGM1,Kallosh-review,BFGM2}.

In particular, $\mathcal{I}$ is quartic\footnote{%
The quartic invariant $\mathcal{I}_{4}$ of $\mathcal{N}=4$ theories was
derived in \cite{CY,CS}.} in charges $Q$ for all rank-three $\mathcal{N}=2$
symmetric spaces \cite{GST}, as well as for $\mathcal{N}=8$ supergravity
(see Eqs. (\ref{I_4-def})-(\ref{Pfaff-democratic}) below). Moreover, since $%
\mathcal{N}\geqslant 3$, $d=4$ supergravities all have symmetric scalar
manifolds, they all admit $W$'s for their various scalar flows, \textit{i.e.}
for each different orbit of the charge vector \cite
{BFGM1,Kallosh-review,BFGM2}.\smallskip

For $\mathcal{N}=8$ supergravity, it follows that
\begin{equation}
W=W\left( \rho _{0},\rho _{1},\rho _{2},\rho _{3},\varphi \right) ,
\label{W-N=8}
\end{equation}
where $\rho _{I}$'s ($I=0,1,2,3$ throughout) are the absolute values of the
skew-eigenvalues of $Z_{AB}$, whose $SU\left( 8\right) $-invariant phase is $%
\varphi $ (see Eq. (\ref{normal-frame}) below). In \cite{DFL-0-brane} the
explicit expressions of $\rho _{I}$'s and $\varphi $ were computed in terms
of the four roots of a quartic algebraic equation, involving the quantities $%
\left( Tr\left( ZZ^{\dag }\right) \right) ^{m+1}$ ($m=0,1,2,3$), as well as
the quartic invariant $\mathcal{I}_{4}$ (see \textit{e.g.} Eqs. (\ref{I_4-1}%
) and (\ref{I_4-2}) below, and also the treatment in \cite{CFMZ-1}).

As shown in \cite{FK-N=8}, two different branches of attractor scalar flows
exist, namely the $\frac{1}{8}$-BPS and the non-BPS branches. Note that $W$
exhibits the same flat directions of $V_{BH}$ at its critical points; such
flat directions span the moduli spaces $\frac{E_{6\left( 2\right) }}{%
SU\left( 6\right) \times SU(2)}$ ($\mathcal{I}_{4}>0$, see Eq. (\ref
{I_4-BPS-large}) below) and $\frac{E_{6\left( 6\right) }}{USp\left( 8\right)
}$ ($\mathcal{I}_{4}<0$, see Eq. (\ref{I_4-nBPS}) below) \cite
{Ferrara-Marrani-2}.

This paper is devoted to the derivation of the $W$'s for both these
branches. This is done by exploiting the ($d=4$) Maurer-Cartan equations of
the exceptional coset $\frac{E_{7\left( 7\right) }}{SU\left( 8\right) }$
(see \textit{e.g.} \cite{ADF-U-duality-revisited} and Refs. therein). We
will show that, while $W_{BPS}$ is given by the highest absolute value of
the skew-eigenvalues of $Z_{AB}$ (consistent with Eq. (\ref{W-BPS-gen})), $%
W_{nBPS}$ is (proportional to) the $USp\left( 8\right) $-singlet of the $%
\mathbf{28}$ of $SU\left( 8\right) $. These results extend to the whole
attractor scalar flow the expression of $W$ which was known for both BPS and
non-BPS attractor solutions after \cite{FK-N=8} (see also \textit{e.g.} \cite
{Ferrara-Marrani-1}). Our investigation and derivation is complementary to
\cite{BMP-1}, where the expression of $W_{nBPS}$ was obtained by making use
of the nilpotent orbits of the $d=3$ geodesic flow obtained through a
timelike reduction (see \textit{e.g.} \cite
{Quantum-Attr-1,Quantum-Attr-2,GLP-1,T-1,BNS-1,T-2,FS-1,FS-2}, and Refs.
therein).\bigskip

The paper is organized as follows.

In Sect. \ref{Democratic-nf} we recall the $SU\left( 6\right) \times
SU\left( 2\right) $-covariant normal frame of $\mathcal{N}=8$ supergravity,
which we dub \textit{``special''} normal frame, and we show that
Maurer-Cartan Eqs. yield a partial differential equation (PDE) for $W$,
whose simplest solution is the BPS superpotential $W_{BPS}$.

Sect. \ref{Undemocratic-nf} is devoted to the analysis of the $USp\left(
8\right) $-covariant normal frame of $\mathcal{N}=8$ supergravity (see
\textit{e.g.} the analysis of \cite{CFGM-1,CFG-1}, and Refs. therein), which
we dub \textit{``symplectic'' }normal frame. We show that in such a normal
frame the Maurer-Cartan Eqs. yield a PDE for $W$, whose simplest solution is
the non-BPS superpotential $W_{nBPS}$. $W$'s are nothing but the singlets in
the decomposition of the $\mathbf{28}$ of $SU\left( 8\right) $ into the
maximal compact subgroup of the stabilizer of the corresponding supporting
charge orbit, \textit{i.e.} respectively into $SU\left( 6\right) \times
SU\left( 2\right) $ (BPS) and $USp\left( 8\right) $ (non-BPS).

Derivations of some relevant formul\ae\ are given in the Appendix, which
concludes the paper.

\section{\label{Democratic-nf}\textit{Special }Normal Frame}

Following \cite{BM,Zumino,FSZ}, through a suitable $SU\left( 8\right) $
transformation the complex skew-symmetric central charge matrix $Z_{AB}$ ($%
A,B=1,...,\mathcal{N}=8$ in the $\mathbf{8}$ of $\mathcal{R}$-symmetry $%
SU\left( 8\right) $) can be \textit{skew-diagonalised}, and thus recast in
\textit{normal} form (see \textit{e.g.} Eq. (87) of \cite{ADOT-1}, adopting
a different convention on the $2\times 2$ symplectic metric $\epsilon $; $%
a=1,2,3$ throughout; unwritten matrix components do vanish throughout):
\begin{eqnarray}
&&Z_{AB}\overset{SU(8)}{\longrightarrow }\left(
\begin{array}{cccc}
z_{0} &  &  &  \\
& z_{1} &  &  \\
&  & z_{2} &  \\
&  &  & z_{3}
\end{array}
\right) \otimes \epsilon =e^{i\frac{\varphi }{4}}\left(
\begin{array}{cccc}
\rho _{0} &  &  &  \\
& \rho _{1} &  &  \\
&  & \rho _{2} &  \\
&  &  & \rho _{3}
\end{array}
\right) \otimes \epsilon ,  \notag \\
\rho _{0},\rho _{a} &\in &\mathbb{R}^{+},~\varphi \in \lbrack 0,8\pi ),
\label{normal-frame}
\end{eqnarray}
where
\begin{equation}
\epsilon \equiv \left(
\begin{array}{cc}
0 & 1 \\
-1 & 0
\end{array}
\right) .  \label{def-epsilon}
\end{equation}
Notice that the second line of Eq. (\ref{normal-frame}) can be obtained from
the first one by performing a suitable $\left( U\left( 1\right) \right) ^{3}$
transformation.

The general definition (\ref{V_BH-gen}) of effective BH potential $V_{BH}$
thus yields
\begin{equation}
V_{BH}=\rho _{0}^{2}+\rho _{1}^{2}+\rho _{2}^{2}+\rho _{3}^{2}.  \label{V_BH}
\end{equation}

Therefore, in the normal frame defined by (\ref{normal-frame}) the
non-vanishing components of $Z_{AB}$ reads as follows:
\begin{eqnarray}
z_{0} &\equiv &Z_{12}=\rho _{0}e^{i\frac{\varphi }{4}};  \label{Z-n-1} \\
z_{1} &\equiv &Z_{34}=\rho _{1}e^{i\frac{\varphi }{4}};  \label{Z-n-2} \\
z_{2} &\equiv &Z_{56}=\rho _{2}e^{i\frac{\varphi }{4}};  \label{Z-n-3} \\
z_{3} &\equiv &Z_{78}=\rho _{3}e^{i\frac{\varphi }{4}}.  \label{Z-n-4}
\end{eqnarray}

Within this parametrization, the (unique) quartic invariant $\mathcal{I}_{4}$
of the $\mathbf{56}$ of the $U$-duality group $E_{7\left( 7\right) }$ (see
\textit{e.g.} \cite{ADF-U-duality-d=4,CFMZ-1}, and Refs. therein)
\begin{eqnarray}
\mathcal{I}_{4} &\equiv &Tr\left( ZZ^{\dag }ZZ^{\dag }\right) -\frac{1}{2^{2}%
}Tr^{2}\left( ZZ^{\dag }\right) +2^{3}Re\left[ \text{Pfaff}\left( Z\right) %
\right] ;  \notag \\
&&  \label{I_4-def} \\
\text{Pfaff}\left( Z\right) &\equiv &\frac{1}{2^{4}4!}\epsilon
^{ABCDEFGH}Z_{AB}Z_{CD}Z_{EF}Z_{GH},  \label{Pfaff-def}
\end{eqnarray}
reads as follows (see \textit{e.g.} \cite{Ferrara-Maldacena}):
\begin{eqnarray}
\mathcal{I}_{4} &=&\sum_{I}\rho _{I}^{4}-2\sum_{I<J}\rho _{I}^{2}\rho
_{J}^{2}+8\rho _{0}\rho _{1}\rho _{2}\rho _{3}\cos \varphi =  \label{I_4-1}
\\
&=&\left( \rho _{0}+\rho _{1}+\rho _{2}+\rho _{3}\right) \left( \rho
_{0}+\rho _{1}-\rho _{2}-\rho _{3}\right) \cdot  \notag \\
&&\cdot \left( \rho _{0}-\rho _{1}+\rho _{2}-\rho _{3}\right) \left( \rho
_{0}-\rho _{1}-\rho _{2}+\rho _{3}\right) +  \notag \\
&&+8\rho _{0}\rho _{1}\rho _{2}\rho _{3}\left( \cos \varphi -1\right) .
\label{I_4-2}
\end{eqnarray}
The \textit{Pfaffian} of $Z_{AB}$, defined by Eq. (\ref{Pfaff-def}), simply
reads
\begin{equation}
\text{Pfaff}\left( Z\right) =Z_{12}Z_{34}Z_{56}Z_{78}=e^{i\varphi
}\prod_{I}\rho _{I}.  \label{Pfaff-democratic}
\end{equation}

It is worth remarking that the skew-diagonal form of $Z_{AB}$ given by Eq. (%
\ref{normal-frame}) is \textit{``democratic'',} in the sense that it fixes
the phases of the four \textit{skew-eigenvalues}
\begin{equation}
z_{I}\equiv \rho _{I}e^{i\varphi _{I}}
\end{equation}
of $Z_{AB}$ to be \textit{all equal:}
\begin{equation}
\varphi _{0}=\varphi _{1}=\varphi _{2}=\varphi _{3}\equiv \frac{\varphi }{4}.
\end{equation}
Actually, this implies some loss of generality, because $SU\left( 8\right) $
only constrains the phases of $z_{I}$'s as follows:
\begin{equation}
\varphi _{0}+\varphi _{1}+\varphi _{2}+\varphi _{3}\equiv \varphi .
\end{equation}
Up to renamings, without loss of generality, the $\left| z_{I}\right| $'s
can be ordered as follows:
\begin{equation}
\rho _{0}\geqslant \rho _{1}\geqslant \rho _{2}\geqslant \rho _{3}.
\label{ordering}
\end{equation}
Notice that $\rho _{I}$'s are $U\left( 8\right) $ invariant, whereas the
overall phase $\varphi $ is invariant under $SU\left( 8\right) $, but not
under $U\left( 8\right) $.

It turns out that the \textit{special} skew-diagonalization (\ref
{normal-frame}) is particularly suitable for the treatment of the $\frac{1}{8%
}$-BPS (\textit{``large'')} attractor flow, as shown in the following
Subsection.

%. Indeed, as we will see below, the $\frac{1}{8}$%
%-BPS (\textit{``large'')} \textit{``fake''} superpotential
%\begin{equation}
%W_{\frac{1}{8}-BPS,large}=\rho _{0}  \label{1/8-BPS-large-W}
%\end{equation}
%can be easily checked to be a solution of the partial differential Eq. (\ref
%{WDE1}) below.

\subsection{\label{Attr-Sols-democratic}Attractor Solutions}

In the \textit{special} normal frame (\ref{normal-frame}), the two attractor
solutions of $\mathcal{N}=8$, $d=4$ supergravity read as follows (see
\textit{e.g.} \cite{FK-N=8}, \cite{CFGM-1}, and Refs. therein; see also the
analysis of \cite{CFG-1} for further detail):

\begin{itemize}
\item  $\frac{1}{8}$-BPS\textit{:}
\begin{eqnarray}
\rho _{0} &\equiv &\rho _{BPS}\in \mathbb{R}_{0}^{+};  \label{1-0} \\
\rho _{1} &=&\rho _{2}=\rho _{3}=0;  \label{l-1} \\
&&\varphi ~\text{undetermined},  \label{l-2}
\end{eqnarray}
thus yielding:
\begin{eqnarray}
Z_{AB,\frac{1}{8}-BPS} &=&e^{i\frac{\varphi }{4}}\rho _{BPS}\left(
\begin{array}{cccc}
1 &  &  &  \\
& 0 &  &  \\
&  & 0 &  \\
&  &  & 0
\end{array}
\right) \otimes \epsilon ;  \label{1/8-BPS-large-horizon-Z} \\
&&  \notag \\
\mathcal{I}_{4}\left( Q_{BPS}\right) &=&\rho _{BPS}^{4}\left( Q_{BPS}\right)
>0,  \label{I_4-BPS-large}
\end{eqnarray}
where \cite{FG-1}
\begin{equation}
Q_{BPS}\in \mathcal{O}_{\frac{1}{8}-BPS,non-deg\text{ }}=\frac{E_{7\left(
7\right) }}{E_{6\left( 2\right) }},
\end{equation}
with maximal compact symmetry $SU\left( 6\right) \times SU\left( 2\right) $.

\item  Non-BPS:
\begin{eqnarray}
\rho _{0} &=&\rho _{1}=\rho _{2}=\rho _{3}\equiv \rho _{nBPS}\in \mathbb{R}%
_{0}^{+};  \label{PA-1} \\
&&\varphi =\pi ,
\end{eqnarray}
thus yielding:
\begin{eqnarray}
Z_{AB,nBPS} &=&e^{i\frac{\pi }{4}}\rho _{nBPS}\Omega _{AB};
\label{nBPS-horizon-Z} \\
&&  \notag \\
\mathcal{I}_{4}\left( Q_{nBPS}\right) &=&-2^{4}\rho _{nBPS}^{4}\left(
Q_{nBPS}\right) <0,  \label{I_4-nBPS}
\end{eqnarray}
where
\begin{equation}
\Omega _{AB}\equiv \left(
\begin{array}{cccc}
1 &  &  &  \\
& 1 &  &  \\
&  & 1 &  \\
&  &  & 1
\end{array}
\right) \otimes \epsilon  \label{Omega}
\end{equation}
is the $8\times 8$ metric of $USp\left( 8\right) $, and \cite{FG-1}
\begin{equation}
Q_{nBPS}\in \mathcal{O}_{nBPS\text{ }}=\frac{E_{7\left( 7\right) }}{%
E_{6\left( 6\right) }},
\end{equation}
with maximal compact symmetry $USp\left( 8\right) $.
\end{itemize}

\subsection{\label{MC-democratic}Maurer-Cartan Equations and PDE for $W$}

Let us now consider the \textit{Maurer-Cartan }Eqs. of $\mathcal{N}=8$, $d=4$
supergravity (see \textit{e.g.} \cite{ADF-U-duality-revisited} and Refs.
therein):
\begin{equation}
\nabla _{i}Z_{AB}=\frac{1}{2}P_{ABCD,i}\overline{Z}^{CD},  \label{MC}
\end{equation}
where the \textit{Vielbein} $1$-form $P_{ABCD}=P_{ABCD,i}d\phi ^{i}$ ($%
i=1,...,70$) of the real homogeneous symmetric scalar manifold
\begin{equation}
M_{\mathcal{N}=8,d=4}=\frac{E_{7\left( 7\right) }}{SU\left( 8\right) }
\end{equation}
sits in the $4$-fold antisymmetric $\mathbf{70}$ of $SU\left( 8\right) $,
and it satisfies the \textit{self-dual reality} condition (see \textit{e.g.}
\cite{ADF-U-duality-d=4})
\begin{equation}
P_{ABCD}=P_{[ABCD]}=\frac{1}{4!}\epsilon _{ABCDEFGH}\overline{P}^{EFGH}.
\label{SDR}
\end{equation}
In order to simplify forthcoming calculations, it is convenient to group $%
SU\left( 8\right) $-indices as follows:
\begin{equation}
12\rightarrow 0;34\rightarrow 1;56\rightarrow 2;78\rightarrow 3.
\label{ind-reduct}
\end{equation}
Thus, for a generic skew-diagonal $Z_{AB}$, \textit{Maurer-Cartan }Eqs. (\ref
{MC}) read
\begin{eqnarray}
\nabla _{i}Z_{0} &=&P_{01,i}\overline{Z}^{1}+P_{02,i}\overline{Z}%
^{2}+P_{03,i}\overline{Z}^{3};  \label{MC-n-1} \\
\nabla _{i}Z_{1} &=&P_{01,i}\overline{Z}^{0}+P_{12,i}\overline{Z}%
^{2}+P_{13,i}\overline{Z}^{3};  \label{MC-n-2} \\
\nabla _{i}Z_{2} &=&P_{02,i}\overline{Z}^{0}+P_{12,i}\overline{Z}%
^{1}+P_{23,i}\overline{Z}^{3};  \label{MC-n-3} \\
\nabla _{i}Z_{3} &=&P_{03,i}\overline{Z}^{0}+P_{13,i}\overline{Z}%
^{1}+P_{23,i}\overline{Z}^{2}.  \label{MC-n-4}
\end{eqnarray}

By disregarding the reality condition (\ref{SDR}) of the \textit{Vielbein} $%
P_{ABCD}$, within the considered \textit{special} normal frame (\ref
{normal-frame}) one can determine the PDE for $W$ in an easy way. Indeed,
Eqs. (\ref{MC}) yield
\begin{eqnarray}
\nabla _{i}\rho _{I} &=&\frac{1}{2}\left( e^{i\varphi /4}\nabla _{i}%
\overline{Z}^{I}+e^{-i\varphi /4}\nabla _{i}{Z}_{J}\right) ;
\label{MC-nnn-2} \\
&&  \notag \\
\nabla _{i}\varphi &=&-2i\nabla _{i}\left( \ln Z_{I}-\ln \overline{Z}%
^{I}\right) =\frac{2}{\rho _{I}}\left( i\,e^{i\varphi /4}\nabla _{i}^{I}%
\overline{Z}-i\,e^{-i\varphi /4}\nabla _{i}Z_{I}\right) .  \notag \\
&&
\end{eqnarray}
Consequently, the total covariant differential of $W$ generally reads (the
sum is expanded in Eq. (\ref{Pp-1-detail}))
\begin{equation}
\nabla _{i}W=\frac{1}{2}\sum_{I<J}\left\{ e^{i\varphi /2}\left( W_{I}\rho
_{J}+W_{J}\rho _{I}\right) +e^{-i\varphi /2}\tilde{\epsilon}^{IJKL}\left(
\overline{W}_{K}\rho _{L}+\overline{W}_{L}\rho _{K}\right) \right\} P_{IJ},
\label{Pp-1}
\end{equation}
%%%%%%%%%%%%%%%%%%%%%%%%%%%%%%%%%%%%%%%%%%%%%%%%%%%%%%%%%%%%%%%%%%%%
where the quantity%
%%%%%%%%%%%%%%%%%%%%%%%%%%%%%%%%%%%%%%%%%%%%%%%%%%%%%%%%%%%%%%%%%%%%
\begin{equation}
W_{I}\equiv \frac{\partial W}{\partial \rho _{I}}+\frac{i}{\rho _{I}}\frac{%
\partial W}{\partial \varphi }\,  \label{W_I-def}
\end{equation}
was introduced.

By performing various steps (detailed in App. A, see Eqs. (\ref{Ps-1})-(\ref
{Ps-2}) therein) and recalling Eq. (\ref{V_BH-W}), the final PDE for the
fake superpotential $W$ reads:%
%%%%%%%%%%%%%%%%%%%%%%%%%%%%%%%%%%%%%%%%%%%%%%%%%%%%%%%%%%%%%%%%%%%%
\begin{eqnarray}
W^{2} &+&\sum_{I,J\neq I}\left\{
\begin{array}{l}
|\left( W_{I}\rho _{J}+W_{J}\rho _{I}\right) |^{2}+ \\
\\
+\frac{1}{2}\left[
\begin{array}{l}
e^{i\varphi }\tilde{\epsilon}^{IJKL}\left( W_{I}\rho _{J}+W_{J}\rho
_{I}\right) \left( W_{K}\rho _{L}+W_{L}\rho _{K}\right) + \\
\\
+e^{-i\varphi }\tilde{\epsilon}^{IJKL}\left( \overline{W}_{I}\rho _{J}+%
\overline{W}_{J}\rho _{I}\right) \left( \overline{W}_{K}\rho _{L}+\overline{W%
}_{L}\rho _{K}\right)
\end{array}
\right]
\end{array}
\right\} {=}  \notag \\
&&  \notag \\
&=&\rho _{0}^{2}+\rho _{1}^{2}+\rho _{2}^{2}+\rho _{3}^{2},
\label{WDE1-impl}
\end{eqnarray}
%%%%%%%%%%%%%%%%%%%%%%%%%%%%%%%%%%%%%%%%%%%%%%%%%%%%%%%%%%%%%%%%%%%%
where all terms of the sum can be found in Eq. (\ref{WDE1}). %
%%%%%%%%%%%%%%%%%%%%%%%%%%%%%%%%%%%%%%%%%%%%%%%%%%%%%%%%%%%%%%%%%%%%

As a consequence of $\mathcal{N}=8$ supersymmetry, Eq. (\ref{WDE1-impl}) is
fully symmetric in $\left\{ \rho _{0},\rho _{1},\rho _{2},\rho _{3}\right\} $%
, and it is straightforward to check that any the $\rho _{I}$'s is a
solution. Following \cite{ADOT-1}, a natural \textit{Ansatz} for $\mathcal{N}%
=8$ solutions is a linear combination of the skew eigenvalues (with constant
coefficients):
\begin{equation}
W=\sum_{I=0}^{3}\,\alpha _{I}\rho _{I}.  \label{Ans-1}
\end{equation}
In this way, the $\frac{1}{8}$-BPS solution reads ($a=1,2,3$)
\begin{equation}
\alpha _{0}=1,\,\alpha _{a}=0.
\end{equation}
Due to the asymptotical meaning of $W$ itself as ADM mass (see Eq. (\ref{ADM}%
)), this corresponds to choosing
\begin{equation}
W_{\frac{1}{8}-BPS}=\rho _{0},  \label{1/8-BPS-large-W}
\end{equation}
namely the highest of the absolute values of the skew-eigenvalues of $Z_{AB}$
as given by ordering (\ref{ordering}), as the solution with the physical
meaning of superpotential.

For non-BPS attractor flow, \textit{Ansatz} (\ref{Ans-1}) produces \cite
{ADOT-1}
\begin{equation}
W_{nBPS}=\frac{1}{2}\left( \rho _{0}+\rho _{1}+\rho _{2}+\rho _{3}\right) ,
\label{Ema-1}
\end{equation}
which however is not the most general one. Indeed, $W_{nBPS}$ given by Eq. (%
\ref{Ema-1}) is a solution \textit{iff} the phase is fixed as
\begin{equation}
\varphi =\pi ,
\end{equation}
thus not describing the most general flow with all five parameters, but
rather a particular case with \textit{double-extremal} phase (see Sect. \ref
{Undemocratic-nf}).

%(see Eq. (\ref{1/8-BPS-large-W})).

%Thus, through a full exploitation of the differential relations given by
%Maurer-Cartan Eqs. (\ref{MC-n-1})-(\ref{MC-n-4}), we re-obtained the result
%of \cite{ADOT-1}, without making use of any \textit{a priori Ansatz}.
Let us notice that the result (\ref{1/8-BPS-large-W}) is an extension to the
whole attractor flow (\textit{i.e.} for all range of the radial coordinate $%
\tau \in \left( -\infty ,0\right] $) of the well-known fact that the
solution of the $\frac{1}{8}$-BPS solution to the $\mathcal{N}=8$ Attractor
Eqs. is obtained by retaining the singlet in the decomposition of $SU\left(
8\right) $ with respect to the stabilizer of the $\frac{1}{8}$-BPS \textit{%
non-degenerate} charge orbit, namely \cite
{ADF-U-duality-d=4,FK-N=8,ADFT-review,Ferrara-Marrani-1}:

\begin{eqnarray}
E_{7\left( 7\right) } &\rightarrow &SU\left( 8\right) \rightarrow SU\left(
6\right) \times SU\left( 2\right) \times U\left( 1\right) ;  \notag \\
&&  \notag \\
\mathbf{56} &\rightarrow &\mathbf{28}+\overline{\mathbf{28}}\rightarrow
\left( \mathbf{15},\mathbf{1}\right) _{+1}+\left( \mathbf{6},\mathbf{2}%
\right) _{-1}+\left( \mathbf{1},\mathbf{1}\right) _{-3}+  \notag \\
&&+\left( \overline{\mathbf{15}},\mathbf{1}\right) _{-1}+\left( \overline{%
\mathbf{6}},\mathbf{2}\right) _{+1}+\left( \mathbf{1},\mathbf{1}\right)
_{+3},  \label{BPS-decomp}
\end{eqnarray}
where the subscripts denote the charge with respect to $U\left( 1\right) $.
The corresponding extension to the whole $\frac{1}{8}$-BPS\textit{\ }%
attractor flow amounts to stating that the superpotential governing the
evolution is given by the ``singlet sector'' $\left( \mathbf{1},\mathbf{1}%
\right) _{+3}+\left( \mathbf{1},\mathbf{1}\right) _{+3}$ in the
decomposition (\ref{BPS-decomp}). In the normal frame (\ref{normal-frame}),
by recalling Eqs. (\ref{Z-n-1})-(\ref{Z-n-4}) and splitting the index of the
$\mathbf{8}$ of $SU\left( 8\right) $ as $A=\widehat{a},\widetilde{a}$, with $%
\widehat{a}=1,2$ and $\widetilde{a}=3,...,8$ (consistently with (\ref
{BPS-decomp})), it then follows that
\begin{equation}
W_{\frac{1}{8}-BPS}=\left| Z_{12}\right| =\rho _{0}.
\end{equation}

\section{\label{Undemocratic-nf}\textit{Symplectic} Normal Frame:\newline
Maurer-Cartan Equations and PDE for $W$}

This Section is devoted to the derivation of the non-BPS \textit{``fake''}
superpotential uniquely from Maurer-Cartan Eqs., with suitable boundary
horizon conditions.

We will obtain $W_{nBPS}$ as a solution of the Maurer-Cartan Eqs. in a
suitably defined manifestly $USp\left( 8\right) $-covariant normal frame
\cite{BMP-1}, in which maximal compact symmetry $USp\left( 8\right) $ of the
non-BPS charge orbit $\frac{E_{7\left( 7\right) }}{E_{6\left( 6\right) }}$
\cite{FG-1} is fully manifest (see \textit{e.g.} also the treatment of \cite
{Ferrara-Marrani-1,CFGM-1,CFG-1}). As it will be evident from subsequent
treatment, such a normal frame is generally and intrinsically not \textit{%
``democratic''} (in the meaning specified at the start of Sect. \ref
{Democratic-nf}).\smallskip

In order to derive the non-BPS \textit{``fake''} superpotential from the
geometric structure encoded in the Maurer-Cartan Eqs., we extend to the
whole attractor flow the well-known fact that the non-BPS solution of the $%
\mathcal{N}=8$ Attractor Eqs. is obtained by retaining the singlet in the
decomposition of $SU\left( 8\right) $ with respect to the stabilizer of the
non-BPS charge orbit, namely \cite{FK-N=8,ADFT-review,Ferrara-Marrani-1}:

\begin{eqnarray}
E_{7\left( 7\right) } &\rightarrow &SU\left( 8\right) \rightarrow USp\left(
8\right) ;  \notag \\
&&  \notag \\
\mathbf{56} &\rightarrow &\mathbf{28}+\overline{\mathbf{28}}\rightarrow
\mathbf{27}+\mathbf{1}+\mathbf{27}^{\prime }+\mathbf{1}^{\prime },
\label{nBPS-decomp}
\end{eqnarray}
where the priming distinguishes the various real irreprs. of $USp\left(
8\right) $, namely the rank-$2$ antisymmetric skew-traceless $\mathbf{27}%
^{\left( \prime \right) }$ and the related skew-trace $\mathbf{1}^{\left(
\prime \right) }$. The corresponding extension to the non-BPS attractor flow
amounts to stating that the superpotential governing the evolution is given
by the ``singlet sector'' $\mathbf{1}+\mathbf{1}^{\prime }$ in the
decomposition (\ref{nBPS-decomp}).

The branching (\ref{nBPS-decomp}) corresponds to decomposing the
skew-diagonal complex matrix $Z_{AB}$ (within the generic normal frame given
by the first line of Eq. (\ref{normal-frame})) into its skew-trace and its
traceless part. This amounts to introducing the following quantities:
\begin{equation}
\left\{
\begin{array}{l}
z_{0}\equiv b+c_{1}+c_{2}+c_{3}; \\
\\
z_{a}\equiv b-c_{a};
\end{array}
\right. \Longleftrightarrow \left\{
\begin{array}{l}
b=\frac{1}{4}\left( z_{0}+\sum_{a}z_{a}\right) ; \\
\\
c_{a}=\frac{1}{4}\left( z_{0}+\sum_{a}z_{a}-4z_{a}\right) ,
\end{array}
\right.  \label{PP-p-1}
\end{equation}
thus yielding
\begin{equation}
Z_{AB}=b~\Omega _{AB}+\mathcal{T}_{0,AB},  \label{ddecomp}
\end{equation}
with $b$ and $\mathcal{T}_{0}$ respectively being (half of) the skew-trace
and the skew-traceless part of the skew-diagonal complex matrix $Z_{AB}$
(within the generic normal frame given by the first line of Eq. (\ref
{normal-frame})):
\begin{eqnarray}
b &\equiv &\frac{1}{8}Z_{AB}\Omega ^{AB};  \label{ddecomp-1} \\
&&  \notag \\
\mathcal{T}_{0,AB} &\equiv &Z_{AB}-\frac{1}{8}Z_{CD}\Omega ^{CD}\Omega
_{AB}=\left(
\begin{array}{cccc}
c_{1}+c_{2}+c_{3} &  &  &  \\
& -c_{1} &  &  \\
&  & -c_{2} &  \\
&  &  & -c_{3}
\end{array}
\right) \otimes \epsilon ,  \notag \\
&&  \label{ddecomp-2}
\end{eqnarray}
where $\Omega _{AB}$ is the $8\times 8$ metric of $USp\left( 8\right) $
defined in (\ref{Omega}).

%%%%%%%%%%%%%%%%%%%%%%%%%%%%%%%%%%%%%%%%%%%%%%%%%%%%%%%%%%%%%%%%%%%%
Following the same steps as in Sect. \ref{Democratic-nf}, with details
explained in App. A (see Eqs. (\ref{bb})-(\ref{P-a-1}) therein), after some
straightforward algebra, one achieves the following result (recall $a=1,2,3$
throughout):
\begin{eqnarray}
&&
\begin{array}{l}
\nabla W\nabla W= \\
\\
=\frac{1}{8}\left\{ \left| 4Re\left[ \left( b\frac{\partial W}{\partial
\overline{b}}-\sum_{a}c_{a}\frac{\partial W}{\partial \overline{c}_{a}}%
\right) +\left( c_{2}+c_{3}\right) \left( -\frac{\partial W}{\partial
\overline{c}_{1}}+\frac{\partial W}{\partial \overline{c}_{2}}+\frac{%
\partial W}{\partial \overline{c}_{3}}\right) \right] +\right. \right. \\
\\
-2iIm\left[ \left( c_{2}+c_{3}\right) \left( \frac{\partial W}{\partial
\overline{b}}-\sum_{a}\frac{\partial W}{\partial \overline{c}_{a}}\right)
+2b\left( -\frac{\partial W}{\partial \overline{c}_{1}}+\frac{\partial W}{%
\partial \overline{c}_{2}}+\frac{\partial W}{\partial \overline{c}_{3}}%
\right) +\right. \\
\\
\quad \left. \left. +2\left( -c_{1}\frac{\partial W}{\partial \overline{c}%
_{1}}+c_{2}\frac{\partial W}{\partial \overline{c}_{2}}+c_{3}\frac{\partial W%
}{\partial \overline{c}_{3}}\right) \right] \right| ^{2}+ \\
\\
\qquad +\left| 4Re\left[ \left( b\frac{\partial W}{\partial \overline{b}}%
-\sum_{a}c_{a}\frac{\partial W}{\partial \overline{c}_{a}}\right) +\left(
c_{1}+c_{3}\right) \left( \frac{\partial W}{\partial \overline{c}_{1}}-\frac{%
\partial W}{\partial \overline{c}_{2}}+\frac{\partial W}{\partial \overline{c%
}_{3}}\right) \right] +\right. \\
\\
-2iIm\left[ \left( c_{1}+c_{3}\right) \left( \frac{\partial W}{\partial
\overline{b}}-\sum_{a}\frac{\partial W}{\partial \overline{c}_{a}}\right)
+2b\left( \frac{\partial W}{\partial \overline{c}_{1}}-\frac{\partial W}{%
\partial \overline{c}_{2}}+\frac{\partial W}{\partial \overline{c}_{3}}%
\right) +\right. \\
\\
\quad \left. \left. +2\left( c_{1}\frac{\partial W}{\partial \overline{c}_{1}%
}-c_{2}\frac{\partial W}{\partial \overline{c}_{2}}+c_{3}\frac{\partial W}{%
\partial \overline{c}_{3}}\right) \right] \right| ^{2}+ \\
\\
+\left| 4Re\left[ \left( b\frac{\partial W}{\partial \overline{b}}%
-\sum_{a}c_{a}\frac{\partial W}{\partial \overline{c}_{a}}\right) +\left(
c_{1}+c_{2}\right) \left( \frac{\partial W}{\partial \overline{c}_{1}}+\frac{%
\partial W}{\partial \overline{c}_{2}}-\frac{\partial W}{\partial \overline{c%
}_{3}}\right) \right] +\right. \\
\\
-2iIm\left[ \left( c_{1}+c_{2}\right) \left( \frac{\partial W}{\partial
\overline{b}}-\sum_{a}\frac{\partial W}{\partial \overline{c}_{a}}\right)
+2b\left( \frac{\partial W}{\partial \overline{c}_{1}}+\frac{\partial W}{%
\partial \overline{c}_{2}}-\frac{\partial W}{\partial \overline{c}_{3}}%
\right) +\right. \\
\\
\left. \qquad \qquad \quad \left. \left. +2\left( c_{1}\frac{\partial W}{%
\partial \overline{c}_{1}}+c_{2}\frac{\partial W}{\partial \overline{c}_{2}}%
-c_{3}\frac{\partial W}{\partial \overline{c}_{3}}\right) \right] \right|
^{2}\right\} .
\end{array}
\notag \\
&&  \label{DWDW}
\end{eqnarray}
%%%%%%%%%%%%%%%%%%%%%%%%%%%%%%%%%%%%%%%%%%%%%%%%%%%%%%%%%%%%%%%%%%%%
%
%
%
%
%

In order to proceed further, group theoretical arguments based on the
reality of the $\mathbf{27}$ and $\mathbf{27}^{\prime }$ of $USp\left(
8\right) $ (see Eq. (\ref{nBPS-decomp})) allow for the following change of
parametrization of the traceless part $\mathcal{T}_{0,AB}$ (see Eq. (\ref
{ddecomp}))
%%%%%%%%%%%%%%%%%%%%%%%%%%%%%%%%%%%%%%%%%%%%%%%%%%%%%%%%%%%%%%%%%%%%
\begin{equation}
c_{a}\equiv \varrho _{\mathbf{27},a}\exp \left( -i\beta \right)
\,\Rightarrow \,\left(
\begin{array}{c}
\frac{\partial }{\partial c_{a}} \\
~ \\
\frac{\partial }{\partial \bar{c}_{a}}
\end{array}
\right) =\left(
\begin{array}{ccc}
e^{i\beta } & ~ & \frac{i}{\xi _{a}}e^{i\beta } \\
~ & ~ & ~ \\
e^{-i\beta } & ~ & -\frac{i}{\xi _{a}}e^{-i\beta }
\end{array}
\right) \left(
\begin{array}{c}
\frac{\partial }{\partial \varrho _{\mathbf{27},a}} \\
~ \\
\frac{\partial }{\partial \beta }
\end{array}
\right) \,,  \label{ccc}
\end{equation}
where, with a slight abuse of language, $\varrho _{\mathbf{27}}$'s generally
denote the degrees of freedom pertaining to the traceless part $\mathcal{T}%
_{0,AB}$ of $Z_{AB}$ (see Eq. (\ref{ddecomp}), and reasoning made above).
%%%%%%%%%%%%%%%%%%%%%%%%%%%%%%%%%%%%%%%%%%%%%%%%%%%%%%%%%%%%%%%%%%%%
Moreover we split the skew-trace into its real and imaginary part
\begin{equation}
b\equiv x+iy\,,\,x,\,y\in \mathbb{R}.
\end{equation}
The reasoning made at the start of the present Section (see Eqs. (\ref
{nBPS-decomp}) and (\ref{ddecomp})) implies the non-BPS \textit{``fake''}
superpotential $W_{nBPS}$ to be related to the skew-trace $b$.

We now proceed by formulating the \textit{Ansatz} that $b$ is independent on
all $\varrho _{\mathbf{27}}$'s introduced in Eq. (\ref{ccc}). As we will see
below, this corresponds to a natural \textit{decoupling Ansatz}\footnote{%
We should also note that this \textit{Ansatz} holds for the particular
solution (\ref{Ema-1}), with $\beta =-\frac{\pi }{4}+2k\pi $ ($k\in \mathbb{Z%
}$) but $\partial W\neq 0$.} for the PDE (\ref{WDE2}) satisfied by $W$,
which will yield to (the) simple(st) solution. This yields the vanishing of
all the derivatives of $W$ with respect to $c_{a}$'s. Thus, Eq. (\ref{DWDW})
reduces to
%%%%%%%%%%%%%%%%%%%%%%%%%%%%%%%%%%%%%%%%%%%%%%%%%%%%%%%%%%%%%%%%%%%%
\begin{eqnarray}
\nabla W\nabla W &=&\frac{1}{8}\left\{
\begin{array}{l}
12\left( x\frac{\partial W}{\partial x}-\frac{\partial W}{\partial y}\right)
+ \\
\\
+\left[ \left( \varrho _{\mathbf{27},1}+\varrho _{\mathbf{27},2}\right)
^{2}+\left( \varrho _{\mathbf{27},1}+\varrho _{\mathbf{27},3}\right)
^{2}+\left( \varrho _{\mathbf{27},2}+\varrho _{\mathbf{27},3}\right) ^{2}%
\right] \cdot \\
\cdot \left( \cos \beta \frac{\partial W}{\partial y}-\sin \beta \frac{%
\partial W}{\partial x}\right) ^{2}
\end{array}
\right\} ,  \notag \\
&&
\end{eqnarray}
%%%%%%%%%%%%%%%%%%%%%%%%%%%%%%%%%%%%%%%%%%%%%%%%%%%%%%%%%%%%%%%%%%%%
so that the whole PDE for the $W$ reads
%%%%%%%%%%%%%%%%%%%%%%%%%%%%%%%%%%%%%%%%%%%%%%%%%%%%%%%%%%%%%%%%%%%%
\begin{eqnarray}
W^{2}+\frac{1}{4}\left\{ 12\left( x\frac{\partial W}{\partial x}-y\frac{%
\partial W}{\partial y}\right) ^{2}+\Delta _{\mathbf{27}}\left( \cos \beta
\frac{\partial W}{\partial y}-\sin \beta \frac{\partial W}{\partial x}%
\right) ^{2}\right\} &=&4\left( x^{2}+y^{2}\right) +\Delta _{\mathbf{27}},
\notag \\
&&  \label{WDE2}
\end{eqnarray}
%%%%%%%%%%%%%%%%%%%%%%%%%%%%%%%%%%%%%%%%%%%%%%%%%%%%%%%%%%%%%%%%%%%%
where the quantity (symmetric in $\left\{ \varrho _{\mathbf{27},1},\varrho _{%
\mathbf{27},2},\varrho _{\mathbf{27},3}\right\} $)
\begin{equation}
\Delta _{\mathbf{27}}\equiv \left( \varrho _{\mathbf{27},1}+\varrho _{%
\mathbf{27},2}\right) ^{2}+\left( \varrho _{\mathbf{27},1}+\varrho _{\mathbf{%
27},3}\right) ^{2}+\left( \varrho _{\mathbf{27},2}+\varrho _{\mathbf{27}%
,3}\right) ^{2}
\end{equation}
was introduced.

Eq. (\ref{WDE2}) is a non-linear PDE in the real functional variables $x$
and $y$. The previous statement that $b$ is independent on all $\varrho _{%
\mathbf{27}}$'s yields that
\begin{equation}
x\neq x\left( \Delta _{\mathbf{27}}\right) ;~~y\neq y\left( \Delta _{\mathbf{%
27}}\right) .  \label{decoupling}
\end{equation}
Under position (\ref{decoupling}), PDE (\ref{WDE2}) naturally \textit{%
decouples }in the following system of PDEs:
\begin{eqnarray}
W^{2}+3\left( x\frac{\partial W}{\partial x}-y\frac{\partial W}{\partial y}%
\right) ^{2} &=&4\left( x^{2}+y^{2}\right) ;  \label{sys-1} \\
\left( \cos \beta \frac{\partial W}{\partial y}-\sin \beta \frac{\partial W}{%
\partial x}\right) ^{2} &=&4.  \label{sys-2}
\end{eqnarray}
%%%%%%%%%%%%%%%%%%%%%%%%%%%%%%%%%%%%%%%%%%%%%%%%%%%%%%%%%%%%%%%%%%%%
PDE (\ref{sys-1}) admits the solution (symmetric in $x$ and $y$)%
%%%%%%%%%%%%%%%%%%%%%%%%%%%%%%%%%%%%%%%%%%%%%%%%%%%%%%%%%%%%%%%%%%%%
\begin{equation}
W\left( x,y\right) =\left( x^{\frac{2}{3}}+y^{\frac{2}{3}}\right) ^{\frac{3}{%
2}},  \label{sol-sys-1}
\end{equation}
%%%%%%%%%%%%%%%%%%%%%%%%%%%%%%%%%%%%%%%%%%%%%%%%%%%%%%%%%%%%%%%%%%%%
which plugged into PDE (\ref{sys-2}) yields the following algebraic equation
for $x$ and $y$ in terms of $\beta $:%
%%%%%%%%%%%%%%%%%%%%%%%%%%%%%%%%%%%%%%%%%%%%%%%%%%%%%%%%%%%%%%%%%%%%
\begin{equation}
\left( x^{2/3}+y^{2/3}\right) \left( x^{1/3}\cos \beta -y^{1/3}\sin \beta
\right) ^{2}=x^{2/3}y^{2/3}\,.  \label{sol-sys-2}
\end{equation}
%%%%%%%%%%%%%%%%%%%%%%%%%%%%%%%%%%%%%%%%%%%%%%%%%%%%%%%%%%%%%%%%%%%%
Eq. (\ref{sol-sys-2}) is in turn solved by (factor $2$ introduced for later
convenience)%
%%%%%%%%%%%%%%%%%%%%%%%%%%%%%%%%%%%%%%%%%%%%%%%%%%%%%%%%%%%%%%%%%%%%
\begin{equation}
x=-2\varrho \sin ^{3}\beta \,,\,y=2\varrho \cos ^{3}\beta \,,
\label{sol-sys-3}
\end{equation}
where $\varrho $ is a real strictly positive number:
\begin{equation}
\varrho \in \mathbb{R}^{+}.  \label{rho}
\end{equation}
In solution (\ref{sol-sys-3}) $\varrho $ is an arbitrary parameter, whose
introduction is possible as a consequence of the homogeneity of degree $0$
of algebraic Eq. (\ref{sol-sys-2}) in $x$ and $y$. In other words, $\varrho $
%parametrizes the dilatation invariance of Eq. (\ref{sol-sys-2}).
can be understood as an integration constant whose meaning has to be
clarified by imposing proper boundary conditions. This is the case for the
requirement of positivity of $\varrho $ which is an asymptotical boundary
condition due to the physical meaning of $W$ that defines the \textit{ADM
mass} $M_{ADM}$ at radial infinity (see Eqs. (\ref{ADM}) and (\ref{WWW})).
%%%%%%%%%%%%%%%%%%%%%%%%%%%%%%%%%%%%%%%%%%%%%%%%%%%%%%%%%%%%%%%%%%%%
Thus, Eqs. (\ref{sol-sys-1}) and (\ref{sol-sys-3}) yield that the final
solution for $W$ reads as follows:
\begin{equation}
W\left( x,y\right) =2\varrho .  \label{WWW}
\end{equation}

By recalling Eqs. (\ref{ddecomp})-(\ref{ddecomp-2}) and (\ref{ccc}), in the
resulting manifestly $USp\left( 8\right) $-covariant normal frame the
central charge matrix $Z_{AB}$ can thus be written as:
\begin{eqnarray}
Z_{AB} &=&2\left( \cos ^{3}\beta +i\sin ^{3}\beta \right) i\,\varrho ~\Omega
_{AB}+  \notag \\
&&  \notag \\
&&+\exp \left( -i\beta \right) \left(
\begin{array}{cccc}
\varrho _{\mathbf{27},1}+\varrho _{\mathbf{27},2}+\varrho _{\mathbf{27},3} &
&  &  \\
& -\varrho _{\mathbf{27},1} &  &  \\
&  & -\varrho _{\mathbf{27},2} &  \\
&  &  & -\varrho _{\mathbf{27},3}
\end{array}
\right) \otimes \epsilon .  \notag \\
&&  \label{Z-undemocratic}
\end{eqnarray}

Eq. (\ref{Z-undemocratic}) determines a parametrization of the \textit{%
symplectic} normal frame (\ref{ddecomp})-(\ref{ddecomp-2}) which is \textit{%
``minimal''}, because it contains only five parameters (see \textit{e.g.}
\cite{Ferrara-Maldacena,FK-N=8}, and Refs. therein) , namely $\left\{ \beta
,\varrho ,\varrho _{\mathbf{27},1},\varrho _{\mathbf{27},2},\varrho _{%
\mathbf{27},3}\right\} $.

In order to consistently characterize solution (\ref{WWW}) as the non-BPS
\textit{``fake''} superpotential, one can use the boundary condition at the
horizon of non-BPS BH. To this end we notice that (see reasoning at the
start of the present Sect.) at non-BPS critical points of $V_{BH,\mathcal{N}%
=8}$ we have
\begin{equation}
\varrho _{\mathbf{27},1}=\varrho _{\mathbf{27},2}=\varrho _{\mathbf{27},3}=0
\label{PA-2}
\end{equation}
so that the parametrization (\ref{Z-undemocratic}) reduces to
\begin{equation}
Z_{AB,nBPS}=2\left( \cos ^{3}\beta _{nBPS}+i\sin ^{3}\beta _{nBPS}\right)
i\,\varrho _{nBPS}\Omega _{AB}.  \label{Z-nBPS-undemocratic}
\end{equation}
This last equation has to be compared with Eq. (\ref{nBPS-horizon-Z}), to
get:
\begin{equation}
2\left( \cos ^{3}\beta _{nBPS}+i\sin ^{3}\beta _{nBPS}\right) i\,\varrho
_{nBPS}=e^{i\frac{\pi }{4}}\rho _{nBPS},
\end{equation}
whose splitting in real and imaginary parts respectively yields:
\begin{eqnarray}
\sqrt{2}\left( \sin ^{3}\beta _{nBPS}-\cos ^{3}\beta _{nBPS}\right) \varrho
_{nBPS} &=&\rho _{nBPS};  \label{Re-1} \\
\cos ^{3}\beta _{nBPS}+\sin ^{3}\beta _{nBPS} &=&0.  \label{Im-1}
\end{eqnarray}
The unique solution of the system (\ref{Re-1})-(\ref{Im-1}) (consistent with
Eq. (\ref{rho})) is found to be
\begin{eqnarray}
\beta _{nBPS} &=&-\frac{\pi }{4}+2k\pi ,~k\in \mathbb{Z}.  \label{A-1} \\
\varrho _{nBPS} &=&\rho _{nBPS}.  \label{A-2}
\end{eqnarray}
In particular Eq. (\ref{A-1}) hints the following redefinition of the
angular parameter:
\begin{equation}
\beta \equiv \alpha -\frac{\pi }{4},
\end{equation}
which yields the parametrization considered in \cite{BMP-1} (up to an
overall factor $\frac{1}{2}$, and with obvious renamings).

The non-BPS nature of the solution (\ref{WWW}) implies the $\mathcal{I}_{4}$
of the $\mathbf{56}$ of $E_{7\left( 7\right) }$ (given by Eqs. (\ref{I_4-1})
and (\ref{I_4-2}) in the \textit{special} normal frame (\ref{normal-frame}))
to be negative. To show this, we rewrite $\mathcal{I}_{4}$ in the manifestly
$USp\left( 8\right) $-covariant parametrization (\ref{Z-undemocratic}),
obtaining \cite{BMP-1}
\begin{equation}
\mathcal{I}_{4}=-2^{4}\sin ^{2}2\beta \left( \varrho \sin 2\beta -\varrho _{%
\mathbf{27},1}{-\varrho _{\mathbf{27},2}-}\varrho _{\mathbf{27},3}\right)
\prod_{a}\left( \varrho \sin 2\beta +\varrho _{\mathbf{27},a}\right) ,
\label{I_4-undemocratic}
\end{equation}
which evaluated at the horizon of non-BPS BH reads:
\begin{equation}
\mathcal{I}_{4,nBPS}=-2^{4}\varrho _{nBPS}^{4}\sin ^{6}\left( 2\beta
_{nBPS}\right) .  \label{I_4-nBPS-undemocratic}
\end{equation}
Using Eqs. (\ref{A-1}) and (\ref{A-2}), Eq. (\ref{I_4-nBPS-undemocratic})
implies
\begin{equation}
\mathcal{I}_{4,nBPS}=-2^{4}\rho _{nBPS}^{4}=-\left. W_{nBPS}^{4}\right|
_{nBPS}<0,
\end{equation}
which confirms the function $W$ given by Eq. (\ref{WWW}) to be the non-BPS
\textit{``fake''} superpotential of $\mathcal{N}=8$, $d=4$ supergravity:
\begin{equation}
W_{nBPS}=2\varrho .  \label{W!!!}
\end{equation}

Thus, $W_{nBPS}$ given by Eq. (\ref{W!!!}) has been proved to be the
simplest solution of the PDE (\ref{WDE2}), determining the non-BPS \textit{%
``fake''} superpotential of $\mathcal{N}=8$, $d=4$ supergravity. The proof
given in the treatment performed above relies completely on the geometric
data encoded into Maurer-Cartan Eqs. (with suitable consistent boundary
horizon conditions), and it is alternative with respect to the treatment
given in \cite{BMP-1}.

As the \textit{special} normal frame (\ref{normal-frame}) has been proved in
Sect. \ref{Democratic-nf} to be more suitable to derive (and describe) $%
\frac{1}{8}$-BPS attractor flow, so the \textit{symplectic} normal frame (%
\ref{Z-undemocratic}) has been proved in this Sect. to be more suitable to
derive (and describe) non-BPS attractor flow.

The expression of $\varrho $ in terms of the five parameters $\left\{ \rho
_{0},\rho _{1},\rho _{2},\rho _{3},\varphi \right\} $ of the \textit{special}
normal frame (\ref{normal-frame}) is not trivial, and it is thoroughly
treated in App. B of \cite{BMP-1}. In general, $\varrho ^{2}$ turns out to
satisfy an algebraic equation of order six with coefficients depending on $%
\left\{ \rho _{0},\rho _{1},\rho _{2},\rho _{3},\varphi \right\} $ and their
(scalar-independent) combination $\mathcal{I}_{4}$, as given by Eq. (B.14)
of \cite{BMP-1} (see also the discussion in \cite{CDFY-2}).

Thus, in general $\varrho ^{2}$ seems not to enjoy an analytical expression.
However, (\textit{at least} one of the) solutions of Eq. (B.14) of \cite
{BMP-1} is a solution of PDE (\ref{WDE1}), yielding $W_{nBPS}$ in the
\textit{special} normal frame (\ref{normal-frame}). Analogously, $W_{\frac{1%
}{8}-BPS}$ given by Eq. (\ref{1/8-BPS-large-W}), suitably translated in the
notation of the \textit{symplectic} normal frame (\ref{Z-undemocratic}) (see
treatment of App. B of \cite{BMP-1}), is a solution of PDE (\ref{WDE2}),
yielding $W_{\frac{1}{8}-BPS}$ in the \textit{symplectic} normal frame (\ref
{Z-undemocratic}). Furthermore, it is here worth mentioning that, through a
suitable re-writing in $\mathcal{N}=2$ language, the results of \cite
{CDFY-1,BMP-1,CDFY-2} are solutions of PDEs (\ref{WDE1}) \textit{and/or} (%
\ref{WDE2}) (eventually through additional reductions to $st^{2}$ or $t^{3}$
models \cite{CDFY-1,BMP-1,CDFY-2}).

\section*{Acknowledgments}

This work is supported in part by the ERC Advanced Grant no. 226455, \textit{%
``Supersymmetry, Quantum Gravity and Gauge Fields''} (\textit{SUPERFIELDS}).

We would like to thank B. L. Cerchiai and R. Kallosh for enlightening
discussions.

S. F. and E. O. would like to thank the \textit{Center for Theoretical
Physics} (SITP) of the University of Stanford, CA USA, and A. M. would like
to thank the Department of Physics and Astronomy, UCLA, CA USA, where part
of this work was done, for kind hospitality and stimulating environment.

The work of S. F.~has been supported in part by D.O.E.~grant
DE-FG03-91ER40662, Task C. The work of A. M. has been supported by an INFN
visiting Theoretical Fellowship at SITP, Stanford University, Stanford, CA
USA. The work of E. O. ~has been supported in part by PRIN Program 2007 of
MIUR. \appendix

\section{Computational Details}

In this Appendix we collect details of the computations determining the
various formul\ae\ of the present paper.

\begin{itemize}
\item  Concerning Sect. \ref{Democratic-nf}, the details are listed below.
%\begin{itemize}
Within the index reduction (\ref{ind-reduct}), the basic multiplication
rules for the vielbein
%%%%%%%%%%%%%%%%%%%%%%%%%%%%%%%%%%%%%%%%%%%%%%%%%%%%%%%%%%%%%%%%%%%%
\begin{eqnarray}
P_{ABCD}\overline{P}^{EFGH} &=&\delta _{ABCD}^{EFGH}\quad ; \\
P_{ABCD}{P}_{EFGH} &=&\epsilon _{ABCDEFGH}\,
\end{eqnarray}
%%%%%%%%%%%%%%%%%%%%%%%%%%%%%%%%%%%%%%%%%%%%%%%%%%%%%%%%%%%%%%%%%%%%
recast as
%%%%%%%%%%%%%%%%%%%%%%%%%%%%%%%%%%%%%%%%%%%%%%%%%%%%%%%%%%%%%%%%%%%%
\begin{eqnarray}
P_{IJ}\bar{P}^{KL} &=&\delta _{I}^{K}\delta _{J}^{L}; \\
P_{IJ}{P}_{KL} &=&\tilde{\epsilon}_{IJKL}\,\equiv \left| \epsilon
_{IJKL}\right| .  \label{j-1}
\end{eqnarray}
%%%%%%%%%%%%%%%%%%%%%%%%%%%%%%%%%%%%%%%%%%%%%%%%%%%%%%%%%%%%%%%%%%%%
Furthermore such rules and Eqs. (\ref{MC}) yield
\begin{eqnarray}
\nabla Z_{I}\nabla Z_{J} &=&\tilde{\epsilon}_{IJKL}\overline{Z}^{K}\overline{%
Z}^{L}; \\
\nabla Z_{I}\nabla \overline{Z}_{J} &=&\delta _{J}^{I}|Z_{I}|=\delta
_{J}^{I}\rho _{I}.  \label{Ps-2}
\end{eqnarray}
Using (\ref{j-1}) and he fully explicited form of Eq. (\ref{Pp-1}) which
reads
\begin{eqnarray}
\nabla _{i}W &=&\frac{1}{2}\left\{ \left[ e^{i\varphi /2}\left( W_{0}\rho
_{1}+W_{1}\rho _{0}\right) +e^{-i\varphi /2}\left( \overline{W}_{2}\rho _{3}+%
\overline{W}_{3}\rho _{2}\right) \right] P_{23}\right.  \notag \\
&&+\left[ e^{i\varphi /2}\left( W_{0}\rho _{2}+W_{2}\rho _{0}\right)
+e^{-i\varphi /2}\left( \overline{W}_{1}\rho _{3}+\overline{W}_{3}\rho
_{1}\right) \right] P_{13}  \notag \\
&&+\left[ e^{i\varphi /2}\left( W_{0}\rho _{3}+W_{3}\rho _{0}\right)
+e^{-i\varphi /2}\left( \overline{W}_{1}\rho _{2}+\overline{W}_{2}\rho
_{1}\right) \right] P_{12}  \notag \\
&&+\left[ e^{i\varphi /2}\left( W_{1}\rho _{2}+W_{2}\rho _{1}\right)
+e^{-i\varphi /2}\left( \overline{W}_{0}\rho _{3}+\overline{W}_{3}\rho
_{0}\right) \right] P_{03}  \notag \\
&&+\left[ e^{i\varphi /2}\left( W_{1}\rho _{3}+W_{3}\rho _{1}\right)
+e^{-i\varphi /2}\left( \overline{W}_{0}\rho _{2}+\overline{W}_{2}\rho
_{0}\right) \right] P_{02}  \notag \\
&&\left. +\left[ e^{i\varphi /2}\left( W_{2}\rho _{3}+W_{3}\rho _{2}\right)
+e^{-i\varphi /2}\left( \overline{W}_{0}\rho _{1}+\overline{W}_{1}\rho
_{0}\right) \right] P_{01}\right\} ,  \notag \\
&&  \label{Pp-1-detail}
\end{eqnarray}
it can be computed that%
%%%%%%%%%%%%%%%%%%%%%%%%%%%%%%%%%%%%%%%%%%%%%%%%%%%%%%%%%%%%%%%%%%%%
\begin{gather}
g^{ij}\nabla _{i}W\nabla _{j}W=\frac{1}{2}\left\{ |\left( W_{0}\rho
_{1}+W_{1}\rho _{0}\right) |^{2}+|\left( W_{0}\rho _{2}+W_{2}\rho
_{0}\right) |^{2}+|\left( W_{0}\rho _{3}+W_{3}\rho _{0}\right) |^{2}+\right.
\notag  \label{Ps-1} \\
\notag \\
\quad \,+\,|\left( W_{1}\rho _{2}+W_{2}\rho _{1}\right) |^{2}+|\left(
W_{1}\rho _{3}+W_{3}\rho _{1}\right) |^{2}+|\left( W_{2}\rho _{3}+W_{3}\rho
_{2}\right) |^{2}+  \notag \\
\notag \\
+\left[ e^{i\varphi }\left( W_{0}\rho _{1}+W_{1}\rho _{0}\right) \left(
W_{2}\rho _{3}+W_{3}\rho _{2}\right) +e^{-i\varphi }\left( \overline{W}%
_{0}\rho _{1}+\overline{W}_{1}\rho _{0}\right) \left( \overline{W}_{2}\rho
_{3}+\overline{W}_{3}\rho _{2}\right) \right] +  \notag \\
\notag \\
+\left[ e^{i\varphi }\left( W_{0}\rho _{2}+W_{2}\rho _{0}\right) \left(
W_{1}\rho _{3}+W_{3}\rho _{1}\right) +e^{-i\varphi }\left( \overline{W}%
_{0}\rho _{2}+\overline{W}_{2}\rho _{0}\right) \left( \overline{W}_{1}\rho
_{3}+\overline{W}_{3}\rho _{1}\right) \right] +  \notag \\
\notag \\
\left. +\left[ e^{i\varphi }\left( W_{0}\rho _{3}+W_{3}\rho _{0}\right)
\left( W_{1}\rho _{2}+W_{2}\rho _{1}\right) +e^{-i\varphi }\left( \overline{W%
}_{0}\rho _{3}+\overline{W}_{3}\rho _{0}\right) \left( \overline{W}_{1}\rho
_{2}+\overline{W}_{2}\rho _{1}\right) \right] \right\} ,  \notag \\
\end{gather}
%%%%%%%%%%%%%%%%%%%%%%%%%%%%%%%%%%%%%%%%%%%%%%%%%%%%%%%%%%%%%%%%%%%%
that, in turns, gives the following expanded form of PDE (\ref{WDE1-impl})
\begin{eqnarray}
&&
\begin{array}{l}
W^{2}+ \\
\\
+\left\{ |\left( W_{0}\rho _{1}+W_{1}\rho _{0}\right) |^{2}+|\left(
W_{0}\rho _{2}+W_{2}\rho _{0}\right) |^{2}+|\left( W_{0}\rho _{3}+W_{3}\rho
_{0}\right) |^{2}+\right. \\
\\
\quad \,\,+|\left( W_{1}\rho _{2}+W_{2}\rho _{1}\right) |^{2}+|\left(
W_{1}\rho _{3}+W_{3}\rho _{1}\right) |^{2}+|\left( W_{2}\rho _{3}+W_{3}\rho
_{2}\right) |^{2}+ \\
\\
+\left[
\begin{array}{l}
e^{i\varphi }\left( W_{0}\rho _{1}+W_{1}\rho _{0}\right) \left( W_{2}\rho
_{3}+W_{3}\rho _{2}\right) + \\
\\
+e^{-i\varphi }\left( \overline{W}_{0}\rho _{1}+\overline{W}_{1}\rho
_{0}\right) \left( \overline{W}_{2}\rho _{3}+\overline{W}_{3}\rho _{2}\right)
\end{array}
\right] + \\
\\
+\left[
\begin{array}{l}
e^{i\varphi }\left( W_{0}\rho _{2}+W_{2}\rho _{0}\right) \left( W_{1}\rho
_{3}+W_{3}\rho _{1}\right) + \\
\\
+e^{-i\varphi }\left( \overline{W}_{0}\rho _{2}+\overline{W}_{2}\rho
_{0}\right) \left( \overline{W}_{1}\rho _{3}+\overline{W}_{3}\rho _{1}\right)
\end{array}
\right] + \\
\\
\left. +\left[
\begin{array}{l}
e^{i\varphi }\left( W_{0}\rho _{3}+W_{3}\rho _{0}\right) \left( W_{1}\rho
_{2}+W_{2}\rho _{1}\right) + \\
\\
+e^{-i\varphi }\left( \overline{W}_{0}\rho _{3}+\overline{W}_{3}\rho
_{0}\right) \left( \overline{W}_{1}\rho _{2}+\overline{W}_{2}\rho _{1}\right)
\end{array}
\right] \right\} = \\
\\
=\rho _{0}^{2}+\rho _{1}^{2}+\rho _{2}^{2}+\rho _{3}^{2}.
\end{array}
\notag \\
&&  \label{WDE1}
\end{eqnarray}
%\end{itemize}

\item  Concerning Sect. \ref{Undemocratic-nf}, the details are listed below.

Within parametrization (\ref{PP-p-1})-(\ref{ddecomp-2}), the Maurer-Cartan
Eqs. (\ref{MC-n-1})-(\ref{MC-n-4}) read as follows:
%%%%%%%%%%%%%%%%%%%%%%%%%%%%%%%%%%%%%%%%%%%%%%%%%%%%%%%%%%%%%%%%%%%%
\begin{eqnarray}
\nabla b &=&\frac{1}{4}\left[
\begin{array}{l}
P_{01}\left( 2\overline{b}+\overline{c}_{2}+\overline{c}_{3}\right)
+P_{02}\left( 2\overline{b}+\overline{c}_{1}+\overline{c}_{3}\right) + \\
\\
+P_{03}\left( 2\overline{b}+\overline{c}_{1}+\overline{c}_{2}\right)
+P_{12}\left( 2\overline{b}-\overline{c}_{1}-\overline{c}_{2}\right) + \\
\\
+P_{13}\left( 2\overline{b}-\overline{c}_{1}-\overline{c}_{3}\right)
+P_{23}\left( 2\overline{b}-\overline{c}_{2}-\overline{c}_{3}\right)
\end{array}
\right] ;  \notag \\
&&  \label{bb} \\
\nabla c_{1} &=&\frac{1}{4}\left[
\begin{array}{l}
P_{01}\left( -2\overline{b}-4\overline{c}_{1}-3\overline{c}_{2}-3\overline{c}%
_{3}\right) +P_{02}\left( 2\overline{b}+\overline{c}_{1}+\overline{c}%
_{3}\right) + \\
\\
+P_{03}\left( 2\overline{b}+\overline{c}_{1}+\overline{c}_{2}\right)
+P_{12}\left( -2\overline{b}-\overline{c}_{1}+3\overline{c}_{2}\right) + \\
\\
+P_{13}\left( -2\overline{b}-\overline{c}_{1}+3\overline{c}_{3}\right)
+P_{23}\left( 2\overline{b}-\overline{c}_{2}-\overline{c}_{3}\right)
\end{array}
\right] ;  \notag \\
&&  \label{c-1} \\
\nabla c_{2} &=&\frac{1}{4}\left[
\begin{array}{l}
P_{01}\left( 2\overline{b}+\overline{c}_{2}+\overline{c}_{3}\right)
+P_{02}\left( -2\overline{b}-3\overline{c}_{1}-4\overline{c}_{2}-3\overline{c%
}_{3}\right) + \\
\\
+P_{03}\left( 2\overline{b}+\overline{c}_{1}+\overline{c}_{2}\right)
+P_{12}\left( -2\overline{b}+3\overline{c}_{1}-\overline{c}_{2}\right) + \\
\\
+P_{13}\left( 2\overline{b}-\overline{c}_{1}-\overline{c}_{3}\right)
+P_{23}\left( -2\overline{b}-\overline{c}_{2}+3\overline{c}_{3}\right)
\end{array}
\right] ;  \notag \\
&&  \label{c-2} \\
\nabla c_{3} &=&\frac{1}{4}\left[
\begin{array}{l}
P_{01}\left( 2\overline{b}+\overline{c}_{2}+\overline{c}_{3}\right)
+P_{02}\left( 2\overline{b}+\overline{c}_{1}+\overline{c}_{3}\right) + \\
\\
+P_{03}\left( -2\overline{b}-3\overline{c}_{1}-3\overline{c}_{2}-4\overline{c%
}_{3}\right) +P_{12}\left( 2\overline{b}-\overline{c}_{1}-\overline{c}%
_{2}\right) + \\
\\
+P_{13}\left( -2\overline{b}+3\overline{c}_{1}-\overline{c}_{3}\right)
+P_{23}\left( -2\overline{b}+3\overline{c}_{2}-\overline{c}_{3}\right)
\end{array}
\right] .  \notag \\
&&  \label{c-3}
\end{eqnarray}
Then, by following the same steps as in Sect. \ref{Democratic-nf}, after
some algebra, one achieves the following result (recall $a=1,2,3$
throughout):
%%%%%%%%%%%%%%%%%%%%%%%%%%%%%%%%%%%%%%%%%%%%%%%%%%%%%%%%%%%%%%%%%%%%
\begin{eqnarray}
&&
\begin{array}{l}
\nabla W= \\
\\
=\frac{1}{4}\left\{ P_{01}\left[
\begin{array}{l}
\left( 2\overline{b}+\overline{c}_{2}+\overline{c}_{3}\right) \left( \frac{%
\partial W}{\partial b}+\sum_{a}\frac{\partial W}{\partial c_{a}}\right) +
\\
+\left( 2b-c_{2}-c_{3}\right) \left( \frac{\partial W}{\partial \overline{b}}%
+\sum_{a}\frac{\partial W}{\partial \overline{c}_{a}}\right) + \\
-4\left( \left( b-c_{2}\right) \frac{\partial W}{\partial \overline{c}_{3}}%
+\left( b-c_{3}\right) \frac{\partial W}{\partial \overline{c}_{2}}\right) +
\\
-4\left( \overline{b}+\overline{c}_{1}+\overline{c}_{2}+\overline{c}%
_{3}\right) \frac{\partial W}{\partial c_{1}}
\end{array}
\right] +\right. \\
\\
+P_{02}\left[
\begin{array}{l}
\left( 2\overline{b}+\overline{c}_{1}+\overline{c}_{3}\right) \left( \frac{%
\partial W}{\partial b}+\sum_{a}\frac{\partial W}{\partial c_{a}}\right) +
\\
+\left( 2b-c_{1}-c_{3}\right) \left( \frac{\partial W}{\partial \overline{b}}%
+\sum_{a}\frac{\partial W}{\partial \overline{c}_{a}}\right) + \\
-4\left( \left( b-c_{1}\right) \frac{\partial W}{\partial \overline{c}_{3}}%
+\left( b-c_{3}\right) \frac{\partial W}{\partial \overline{c}_{1}}\right) +
\\
-4\left( \overline{b}+\overline{c}_{1}+\overline{c}_{2}+\overline{c}%
_{3}\right) \frac{\partial W}{\partial c_{2}}
\end{array}
\right] + \\
\\
+P_{03}\left[
\begin{array}{l}
\left( 2\overline{b}+\overline{c}_{1}+\overline{c}_{2}\right) \left( \frac{%
\partial W}{\partial b}+\sum_{a}\frac{\partial W}{\partial c_{a}}\right) +
\\
+\left( 2b-c_{1}-c_{3}\right) \left( \frac{\partial W}{\partial \overline{b}}%
+\sum_{a}\frac{\partial W}{\partial \overline{c}_{a}}\right) + \\
-4\left( \left( b-c_{1}\right) \frac{\partial W}{\partial \overline{c}_{2}}%
+\left( b-c_{2}\right) \frac{\partial W}{\partial \overline{c}_{1}}\right) +
\\
-4\left( \overline{b}+\overline{c}_{1}+\overline{c}_{2}+\overline{c}%
_{3}\right) \frac{\partial W}{\partial c_{3}}
\end{array}
\right. + \\
\\
+P_{12}\left[
\begin{array}{l}
\left( 2\overline{b}-\overline{c}_{1}-\overline{c}_{2}\right) \left( \frac{%
\partial W}{\partial b}+\sum_{a}\frac{\partial W}{\partial c_{a}}\right) +
\\
+\left( 2b+c_{1}+c_{3}\right) \left( \frac{\partial W}{\partial \overline{b}}%
+\sum_{a}\frac{\partial W}{\partial \overline{c}_{a}}\right) + \\
-4\left( \left( \overline{b}-\overline{c}_{1}\right) \frac{\partial W}{%
\partial c_{2}}+\left( \overline{b}-\overline{c}_{2}\right) \frac{\partial W%
}{\partial c_{1}}\right) + \\
-4\left( b+c_{1}+c_{2}+c_{3}\right) \frac{\partial W}{\partial \overline{c}%
_{3}}
\end{array}
\right] + \\
\\
+P_{13}\left[
\begin{array}{l}
\left( 2\overline{b}-\overline{c}_{1}-\overline{c}_{3}\right) \left( \frac{%
\partial W}{\partial b}+\sum_{a}\frac{\partial W}{\partial c_{a}}\right) +
\\
+\left( 2b+c_{1}+c_{3}\right) \left( \frac{\partial W}{\partial \overline{b}}%
+\sum_{a}\frac{\partial W}{\partial \overline{c}_{a}}\right) + \\
-4\left( \left( \overline{b}-\overline{c}_{1}\right) \frac{\partial W}{%
\partial c_{3}}+\left( \overline{b}-\overline{c}_{3}\right) \frac{\partial W%
}{\partial c_{1}}\right) + \\
-4\left( b+c_{1}+c_{2}+c_{3}\right) \frac{\partial W}{\partial \overline{c}%
_{2}}
\end{array}
\right] + \\
\\
\qquad \left. +P_{23}\left[
\begin{array}{l}
\left( 2\overline{b}-\overline{c}_{2}-\overline{c}_{3}\right) \left( \frac{%
\partial W}{\partial b}+\sum_{a}\frac{\partial W}{\partial c_{a}}\right) +
\\
+\left( 2b+c_{2}+c_{3}\right) \left( \frac{\partial W}{\partial \overline{b}}%
+\sum_{a}\frac{\partial W}{\partial \overline{c}_{a}}\right) + \\
-4\left( \left( \overline{b}-\overline{c}_{2}\right) \frac{\partial W}{%
\partial c_{3}}+\left( \overline{b}-\overline{c}_{3}\right) \frac{\partial W%
}{\partial c_{2}}\right) + \\
-4\left( b+c_{1}+c_{2}+c_{3}\right) \frac{\partial W}{\partial \overline{c}%
_{1}}
\end{array}
\right] \right\} .
\end{array}
\notag \\
&&  \label{P-a-1}
\end{eqnarray}
%%%%%%%%%%%%%%%%%%%%%%%%%%%%%%%%%%%%%%%%%%%%%%%%%%%%%%%%%%%%%%%%%%%%
It is worth noticing that the coefficient of the vielbein $P_{IJ}$ (recall $%
I=0,1,2,3$ throughout) is the complex conjugate of the coefficient of $%
P_{KL} $, with $K,L\neq I,J$. In other words, in order to compute the term $%
\nabla W\nabla W$ one has just to sum up the squares of the real and
imaginary part of each coefficient, thus obtaining Eq. (\ref{DWDW}).

\end{itemize}

\end{document}